\documentclass[final,10pt,twocolumn,authoryear]{elsarticle}
\usepackage[utf8]{inputenc} % To have special characters
\usepackage{lineno}
\usepackage[breaklinks=true,
colorlinks=true,
urlcolor=blue]{hyperref}
\usepackage{color}
\modulolinenumbers[5]

\journal{ArXiv}
\begin{document}
\begin{frontmatter}
%%%%%%%%%%%%%%%%%%%%%% TITLE %%%%%%%%%%%%%%%%%%%%%%%%
\title{CARE-rCortex: a Matlab toolbox for the analysis of CArdio-REspiratory-related activity in the Cortex}

%%%%%%%%%%%%%%%%%%%%%% AUTHORS %%%%%%%%%%%%%%%%%%%%%%%%
\author[a,b]{F. Grosselin\corref{mycorrespondingauthor}}
\cortext[mycorrespondingauthor]{Corresponding author: grosselin.fanny@gmail.com}
%\ead{grosselin.fanny@gmail.com}
\author[b,c]{X. Navarro-Sune}
\author[c,d]{M. Raux}
\author[c,e]{T. Similowski}
\author[f]{M. Chavez}
\address[a]{Sorbonne Universit\'es, UPMC Univ. Paris 06, INSERM U-1127, CNRS UMR-7225, Institut du Cerveau et de la Moelle
\'Epini\`ere (ICM), Groupe Hospitalier Piti\'e Salp\^etri\`ere-Charles Foix, 75013, Paris, France}
\address[b]{myBrainTechnologies, 75010, Paris, France}
\address[c]{Sorbonne Universit\'es, UPMC Univ. Paris 06, INSERM UMRS1158, Neurophysiologie Respiratoire Exp\'erimentale et Clinique,  75005, Paris, France}
\address[d]{AP-HP, Groupe Hospitalier Piti\'e Salp\^etri\`ere-Charles Foix, D\'epartement 
d’Anesth\'esie-R\'eanimation, 75013, Paris, France}
\address[e]{AP-HP, Groupe Hospitalier Piti\'e Salp\^etri\`ere-Charles Foix, Service de Pneumologie et R\'eanimation M\'edicale du D\'epartement R3S, 75013, Paris, France}
\address[f]{CNRS UMR-7225, Groupe Hospitalier Piti\'e-Salp\^etri\`ere-Charles Foix, 75013, Paris, France}

%%%%%%%%%%%%%%%%%%%%%% ABSTRACT %%%%%%%%%%%%%%%%%%%%%%%%
\begin{abstract}
\textit{Background:} Although cardio-respiratory (CR) system is generally controlled by the autonomic nervous system, interactions between the cortex and these primary functions are receiving an increasing interest in neurosciences. 
\textit{New method:} In general, the timing of such internally paced events (e.g. heartbeats or respiratory cycles) may display a large variability. For the analysis of such CR event-related EEG potentials, a baseline must be correctly associated to each cycle of detected events. The open-source toolbox CARE-rCortex provides an easy-to-use interface to detect CR events, define baselines, and analyse in time-frequency (TF) domain the CR-based EEG potentials. 
\textit{Results:} CARE-rCortex provides some practical tools to detect and validate these CR events. Users can define baselines time-locked to a phase of respiratory or heart cycle. A statistical test has also been integrated to highlight significant points of the TF maps with respect to the baseline. We illustrate the use of CARE-rCortex with the analysis of two real cardio-respiratory datasets.
\textit{Comparison with existing methods:} Compared to other open-source toolboxes, CARE-rCortex allows users to automatically detect CR events, to define and check baselines for each detected event. Different baseline normalizations can be used in the TF analysis of EEG epochs. 
\textit{Conclusions:} The analysis of CR-related EEG activities could provide valuable information about cognitive or pathological brain states. CARE-rCortex runs in Matlab as a plug-in of the EEGLAB software, and it is publicly available at \url{https://github.com/FannyGrosselin/CARE-rCortex}.
\end{abstract}

%%%%%%%%%%%%%%%%%%%%%% KEYWORDS %%%%%%%%%%%%%%%%%%%%%%%%
\end{frontmatter}

%\linenumbers
%%%%%%%%%%%%%%%%%%%%%% INTRODUCTION %%%%%%%%%%%%%%%%%%%%%%%%
\section{Introduction}
\begin{figure}[t!]
\centering %from left, bottom, right and top
\includegraphics[width=0.9\columnwidth]{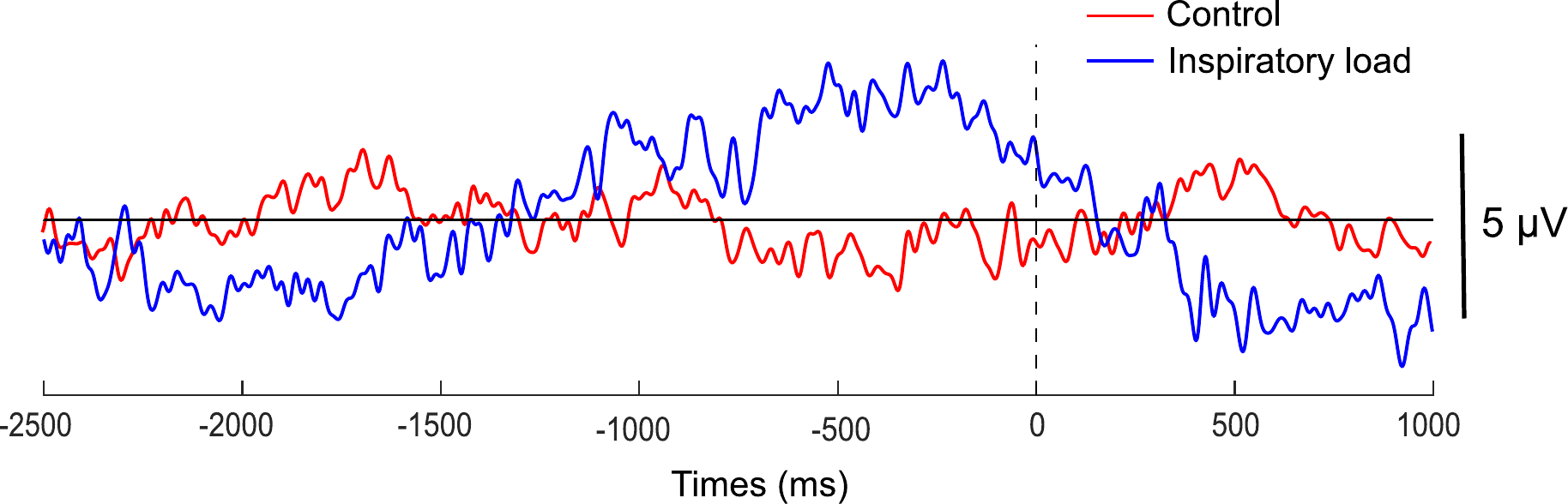}
\caption{Example of respiratory event-related potential. Averaged amplitude across the epochs of the central EEG activity (Cz electrode) of a subject during spontaneously breathing (in red) or during an inspiratory load task (in blue). The dashed vertical line indicates the inspiration onset.}
\label{respiratory_ERP}
\end{figure}

Interactions between brain and cardiac or respiratory activity is receiving an increasing interest in neurosciences~\citep{dubois_neurophysiological_2016,perakakis_neural_2017}. These primary functions are controlled by the autonomic nervous system, particularly the medulla oblongata and the pons \citep{bianchi_central_1995} for the respiration. The medulla oblongata also plays an important role in cardiac activities with ganglia along the cervical and thoracic spinal cord \citep{shen_role_2014}.
However, specific events can modulate the cardio-respiratory (CR) system through the cortex. 
In humans, the breathing function can be modulated by emotions or any external constraint limiting the airflow (e.g. desynchronization between a patient's breathing and the airflow delivered by a ventilator, or breathing against an inspiratory load as illustrated in Fig.~\ref{respiratory_ERP}). In this case, the control of breathing comes from suprapontine structures, especially from cortical structures \citep{colebatch_regional_1991, ramsay_regional_1993, straus_comment_2005}. 

Changes in breathing rhythms are captured by somatosensory mechanoreceptors in the chest and are related to an activation of the cortex like other mechanoreceptors located in different locations in the body. These afferent signals impact the ongoing efferent drive which regulates cardio-respiratory function through a nested hierarchy of simple reflexes and more complex higher level feedback systems \citep{dirlich_topography_1998}.

\begin{figure}[t!]
\centering %from left, bottom, right and top
\includegraphics[width=0.9\columnwidth]{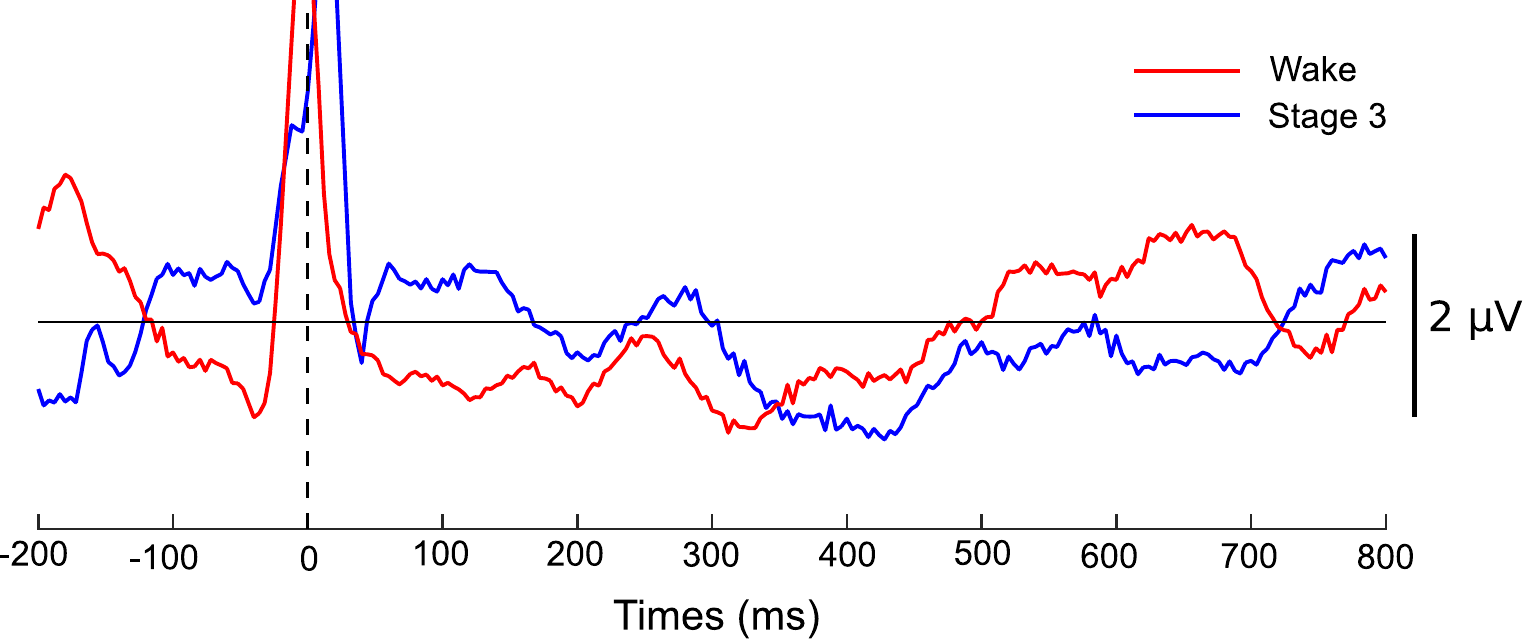}
\caption{Example of heartbeat-related potential. Averaged amplitude of EEG signals at channel C3 in a healthy awake subject during resting state with closed eyes (in red), and during stage 3 of the sleep (in blue). The dashed line indicates the R-peak time.}
\label{cardiac_potential}
\end{figure}

Different works have also shown an effect of these somatosensory signals on the EEG with experimental studies based on heartbeat event-related potentials. Indeed, several studies had highlighted an EEG potential about 300-600ms after the R-wave of the heart especially during an attentional task of focusing on the heartbeats \citep{montoya_heartbeat_1993,schandry_event-related_1996, dirlich_topography_1998} (see Fig.~\ref{cardiac_potential}). This potential related to afferent cardiac information reflects the awareness of one's own heartbeats \citep{gray_cortical_2007}.

All these research studies are based on evoked related potentials (ERP), i.e. an estimate of the response that is time-locked to a stimulus. Event-related changes at different EEG frequency bands constitute an important indicator of the underlying brain processes. Sensory processing and motor behaviour are connected, for instance, with the localized decrease of power in certain frequency bands \citep{durka_statistical_2004}. Therefore, the quantification of significant changes in the EEG time-frequency content associated to the externally/internally paced events needs a normalization with an appropriate baseline (e.g. the baseline time period should always end before each trial onset). 

As mentioned above, the timing of internally paced events (e.g. heartbeats or respiratory cycles) can be modulated by different internal mechanisms, hence highly variable even within a subject. To study the significance of cardio-respiratory related EEG changes, researchers need analysis tools to: \textit{i)} automatically detect non-regular events and, \textit{ii)} define a baseline correctly locked to each detected event.

EEGLAB \citep{delorme_eeglab:_2004} is a well-known interactive Matlab (The Mathworks, Inc., MA) toolbox for processing electrophysiological data. It proposes different easy-to-use tools for single channel time-frequency (TF) representation, such as a baseline-normalized spectrogram or the analysis of event-related power changes \citep{makeig_auditory_1993}. 
FieldTrip \citep{oostenveld_fieldtrip:_2010} is another Matlab based toolbox that also provides TF analysis by applying a normalization with respect to a fixed baseline interval. ERPWAVELAB \citep{morup_erpwavelab:_2007} also proposes a tool to analyse event-related potentials in both EEG and magneto-encephalography (MEG) data in TF domain. As additional features, ERPWAVELAB offers the possibility of performing multi-subject and multiple condition analysis and provides tools for artefact rejection in TF domain.

However, all these tools do not provide event detection from cardio-respiratory signals and do not lock baseline to time-variable events.
HEPLAB \citep{pandelis_perakakis_2018_1164232} was recently introduced to facilitate heartbeat evoked potential (HEP) analysis. It computes heartbeat-evoked events in EEGLAB from continuous EEG signals using R events. These events are automatically detected by means of an external toolbox, ECGLAB \citep{vicente_ecglab:_2013}. FieldTrip is then recommended for statistical analysis between groups and conditions. Another Matlab based toolbox, ECG-kit \citep{demski_ecg-kit:_2016}, includes several algorithms to detect heartbeats and pulse, classifies cardiac events and provides an ECG delineator. However, these tools are limited to the analysis of ECG activities in time domain.

In this work, we propose a new toolbox, CARE-rCortex, that includes useful features to analyse the EEG potentials related to cardiac or breathing rhythms. It proposes an automated \emph{detection} and \emph{validation} of CR events such as the heartbeats, and the breathing cycles. Using detected events, users can define a time-locked optimal baseline. CARE-rCortex also integrates a \emph{manual validation of baseline} location to correct possible misplacements. Moreover, TF maps are purposely designed for frequency bands of interest, and can be normalized by the baseline time-locked to the chosen CR events.

The paper is structured as follows. Firstly, an overview of the main functionalities of our toolbox is presented. This is followed by a brief description of the methods implemented in the toolbox (event detection, baseline management, time-frequency maps and significance analysis). Finally, we illustrate the usefulness of CARE-rCortex with the analysis of two real cardio-respiratory datasets. 

%%%%%%%%%%%%%%%%%%%%%%%%% METHODS %%%%%%%%%%%%%%%%%%%%%%%%%%
\section{Methods}
\subsection{Toolbox overview}
CARE-rCortex was designed as a plug-in for EEGLAB software using Matlab graphical user interface (GUI). The GUI allows users to analyse data interactively without Matlab programming experience, although experienced users can also run CARE-rCortex functions from the command line. To start using our toolbox, EEG or other electrophysiological data have to be firstly loaded into EEGLAB workspace. Depending on the data format, some supplementary packages could be required by EEGLAB.

\begin{figure*}[ht!]
\centering %from left, bottom, right and top
\includegraphics[width=0.7\textwidth, height=0.95\textheight]{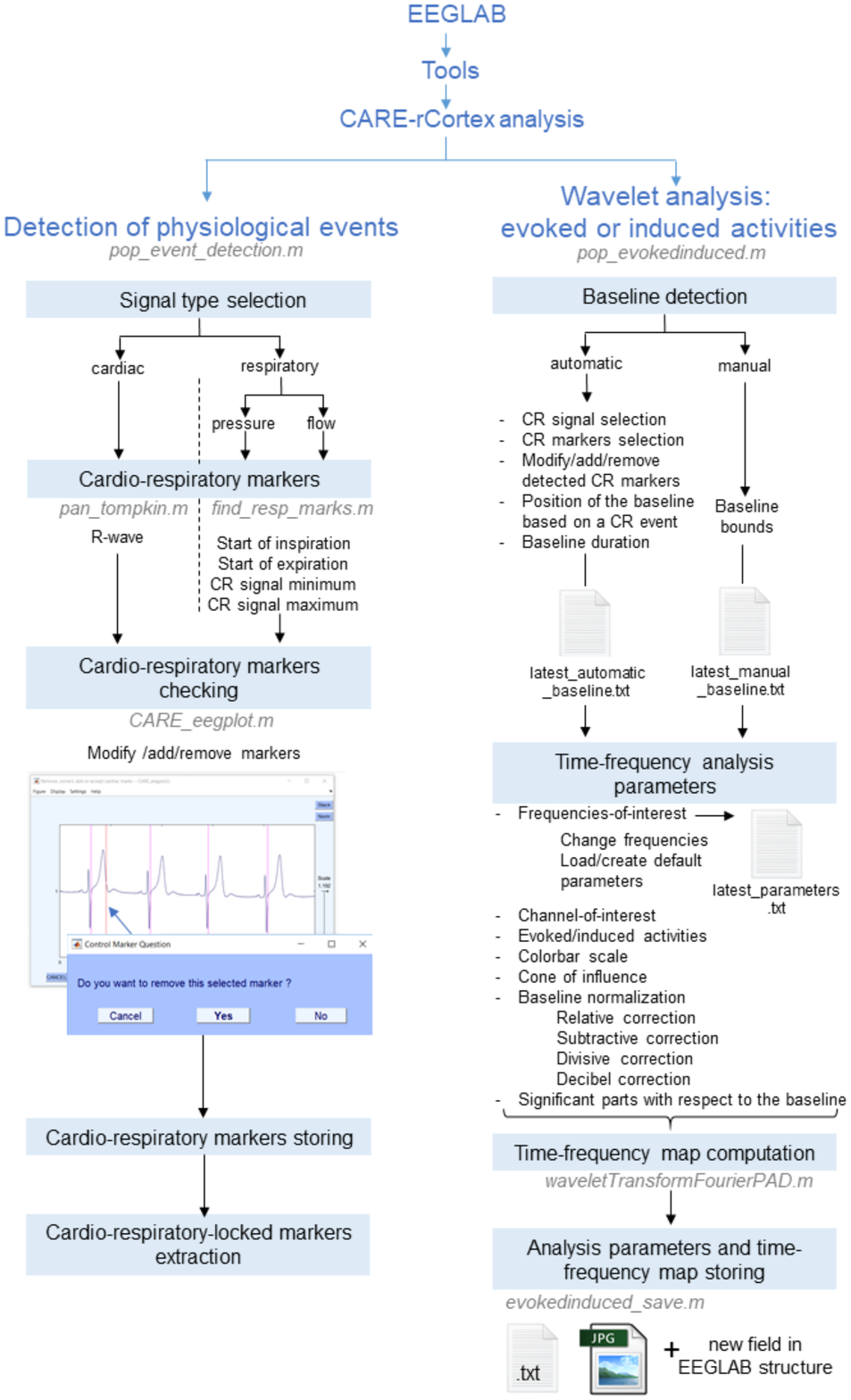}
\caption{Overview of the functionalities of CARE-rCortex}
\label{toolbox_overview}
\end{figure*}

\subsubsection{Detection of events and creation of markers}
After loading the data, EEG epochs have to be extracted with respect to a CR event. These markers are generally selected as time-locking event type(s) during the extraction of epochs by EEGLAB. Because the markers are not always related to CR recordings, CARE-rCortex provides a tool to detect CR events and create the corresponding markers (Fig. \ref{toolbox_overview}). Namely, our toolbox allows the detection of: \textit{i)} the R-wave of the QRS complex in the ECG signals; or \textit{ii)} different respiratory events such as the inspiration or expiration onsets, the maxima or minima of the respiratory pressure or airflow signals. Before saving them as markers for EEGLAB, a manual validation of the detected events can be done. During this validation, events can be added, modified or deleted. 

\subsubsection{Artefact reduction}
EEG recordings are often contaminated by non-neural physiological activities, as well as other external or environmental noises, that seriously degrade the signals of interest \citep{uriguen_2015}.  Such perturbations include classical eye movements and muscular-related activities, but also breathing or heart cycle-related artefacts. The heart's electrical activity for instance, can be measured anywhere on the body surface and can also be observed in EEG recordings along with brain electric components. This inherent artefact, termed cardiac field artefact (CFA), is most prominent during ventricular depolarization (QRS complex), and it is characterized by a sharp potential synchronized to the R-peak, corresponding to the onset of ventricular contraction \citep{dirlich_cfa_1997, dirlich_topography_1998}.

Since the user can take advantage of the built-in EEGLAB functions or other external plug-ins for artefact reduction, CARE-rCortex does not provide further tools for such purpose. For instance,  ``Reject data epochs''  or ``Reject continuous data'' functions can be used to reject EEG segments contaminated with muscle, ocular or head movements-related artefacts. Spatial filters or independent component analysis (ICA) can effectively remove interferences from a wide variety of artefactual sources in EEG recordings, including CFA, but only when the number of channels and the amount of data are large enough \citep{uriguen_2015}. 

For epoched EEG analysis, different template subtraction methods can also be used to reduce the CFA~\citep{Debener_2009}. Nevertheless, such methods assume a lack of correlation between neural activity and cardiac activity and a temporal stability of the artefact, which may result in inaccurate artefact estimation and therefore lead to greater residual contamination after subtraction \citep{Debener_2009}.

\subsubsection{Baseline management}
\begin{figure}[h!]
\centering %from left, bottom, right and top
\includegraphics[width=0.5\textwidth]{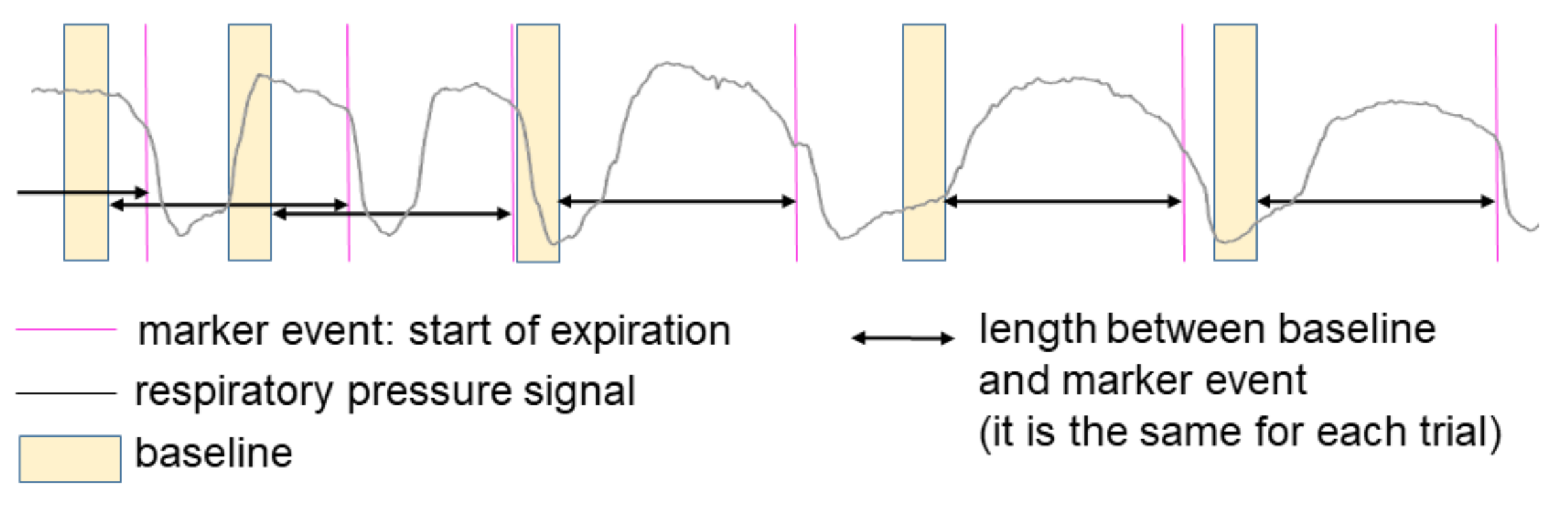}
\caption{Example of improper baseline selection for a respiratory pressure signal. To analyse pre-expiratory activity, baselines should be placed in the pre-inspiratory period of the pressure cycle (as it is the case for the 4$^{th}$ baseline). In this example, some baselines are either located in the previous cycle or out of the pre-inspiratory period (1$^{st}$, 2$^{nd}$ and 3$^{rd}$ baselines).}
\label{bad_baselines}
\end{figure}

To analyse CR event-related potentials, a baseline is usually removed from EEG epochs. The baseline term refers to the basal EEG activity when no actions are being prepared, and is usually selected from a period sufficiently far from the onset of the event of interest. To analyse pre-expiration potentials, for instance, a baseline should be chosen at the beginning of the inspiration time. In most of the studies, the baseline is the same for each epoch and it is characterized by its length and its position from the marker event. Nevertheless, when CR events are studied, some baseline periods may overlap the targeted activity in previous or posterior cycles (see an example in Fig. \ref{bad_baselines}). 

Our toolbox allows researchers to define and correct the baseline of CR events through a GUI interface (Fig. \ref{CARE-rCortex_GUI}). This can be done either manually or automatically according to the detected CR cycles. To accurately define the baseline of detected events, CARE-rCortex displays a histogram of time distances between consecutive markers. Baseline lengths and their positions can be thus chosen as a function of the distances between CR events. The automatic setting of baselines can be easily checked with an interactive plot, allowing the user to remove or add baselines by simply clicking on the plot. Of note, epochs must have only one baseline, located between its start and the event marker. Otherwise, a warning message suggests solutions to fix the problem. 

\subsubsection{Time-frequency maps}
The parametrization of the TF analysis can be done within the same GUI interface. Researchers can select the frequency range of interest for the analysis as well as the frequency resolution. By default, the minimal frequency resolution is set at $0.25$~Hz, and the lowest frequency at $2$~Hz, for a fast and optimal EEG potential analysis related to CR events.

\begin{figure*}[t!]
\centering %from left, bottom, right and top
\includegraphics[width=\textwidth, height=0.4\textheight]{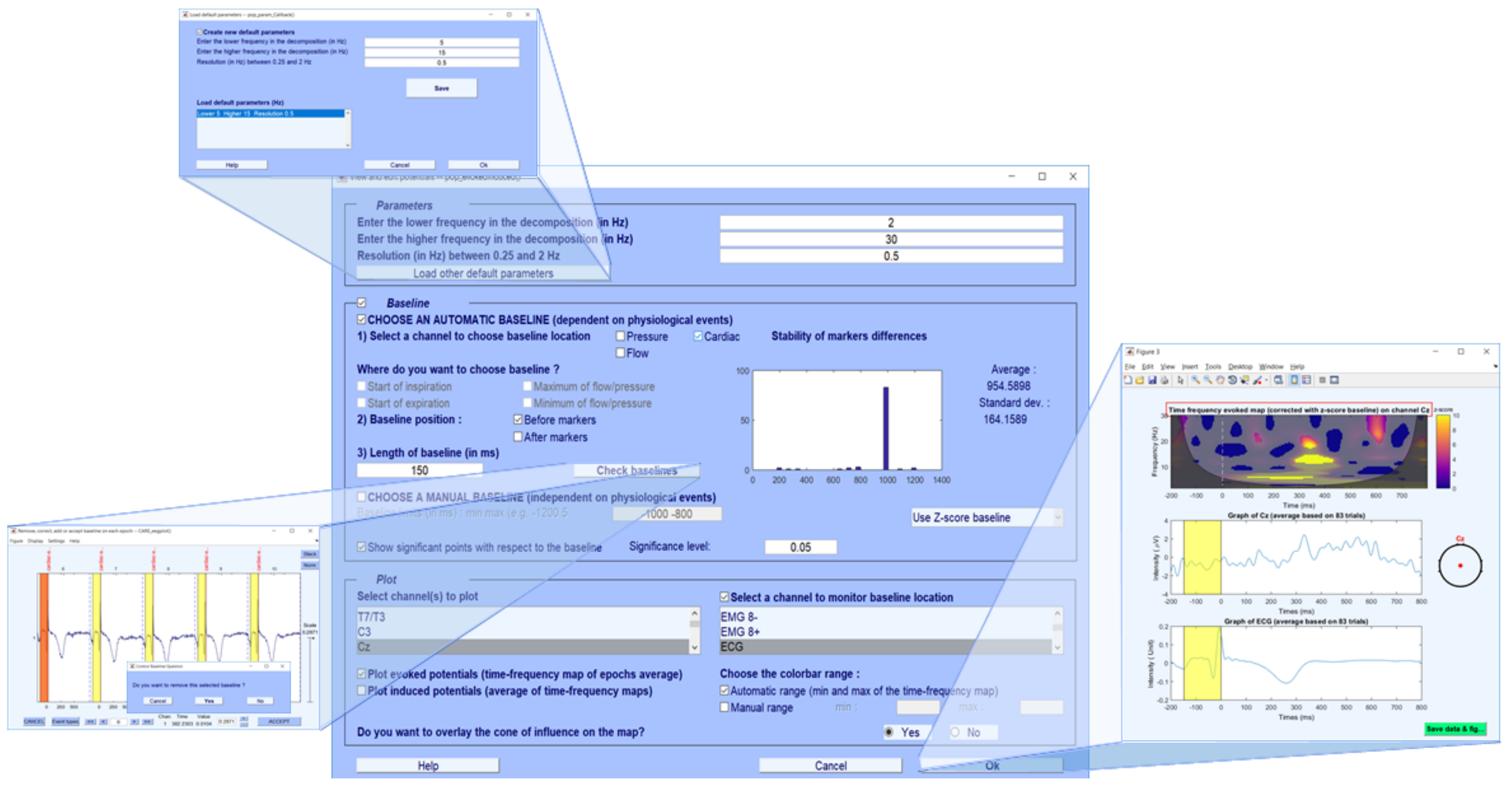}
\caption{The Graphical User Interface of the CARE-rCortex toolbox.}
\label{CARE-rCortex_GUI}
\end{figure*}

The GUI of CARE-rCortex allows researchers to plot the result of time-frequency analysis (Fig. \ref{toolbox_overview} and \ref{CARE-rCortex_GUI}). For a given EEG channel, the time-frequency analysis can be performed in two different ways \citep{tallon-baudry_oscillatory_1999}: either time-frequency map of epoch average (evoked activities) or the average of time-frequency maps (induced activities).
For each TF map, a \textit{cone of influence} can be also displayed to delimit the TF region not influenced by edge effects \citep{torrence_practical_1998}. For the selected channels, CARE-rCortex can also display the EEG epoch average with the median baseline region indicated on the plot. 

Different parameters of the time-frequency analysis, baseline location and length can be loaded or saved in external configuration files. CARE-rCortex will load these values in forthcoming user sessions. Users can also export figures containing time-frequency maps with the results. Finally, time-frequency data and the corresponding significant masks are automatically saved into a new EEGLAB data structure field.

\subsection{Methods implemented in the analysis}
\subsubsection{Cardio-respiratory event detection}
To detect cardiac events (R peaks in QRS complex), we adopted Pan-Tompkins's algorithm~\citep{pan_real-time_1985}, using the Matlab implementation done by \citep{sedghamiz_matlab_2014}. This is a real-time algorithm that reliably recognizes QRS complex based on the analysis of slope, amplitude and width. The algorithm automatically adjusts thresholds and parameters to be adapted to ECG changes such as QRS morphology and heart rate \citep{pan_real-time_1985}. Breathing cycles are detected using a peak detector, adapting the algorithm described in \citep{navarro_resp_2015} to both air flow and pressure signals. 

\subsubsection{Baseline management}
A baseline period is associated to each CR event detected by the previously mentioned algorithms. The baseline length can be chosen by the user. To check that each baseline is placed before or after a given event, we implemented a functionality to visualize and manage the baselines. This function was inspired from \textit{eegplot}, an EEGLAB script to visualize and check the EEG signal in each epoch.

\subsubsection{Time-frequency analysis}
\paragraph{Morlet wavelet decomposition}
The wavelet transform (WT) is a useful analysis tool for time-frequency representation of non-stationary signals, obtained by convolving the signal \textit{s} with a scaled and translated wavelet function \citep{mallat_book_1998}. The WT provides thus a decomposition of the signal variance at different scales. Although continuous wavelets often yield a redundant decomposition of analysed signals (the information extracted from a given scale band slightly overlaps with that from neighbouring scales), they are more robust to noise as compared with other decomposition schemes \citep{mallat_book_1998, torrence_practical_1998}. The complex Morlet wavelet decomposition is employed in the toolbox as it
provides a very good time localization with a high frequency resolution. In addition, the Morlet wavelet has the advantage of having both real and imaginary parts, which allows a separation of the phase and the amplitude of the studied signal \citep{schiff_et_al_1994,van_vugt_2007}.

\paragraph{Cone of influence}
For finite-length time series, numerical errors in the wavelet spectrum will occur at the beginning and at the end of the segment. The area of the TF map where such effects are relevant, the so-called \textit{cone of influence}, is chosen here as the e-folding time (the wavelet power for a discontinuity at the edges drops by a factor $e^{-2}$) of the Morlet wavelet function  \citep{torrence_practical_1998}.
 
\paragraph{Two TF representations}
CARE-rCortex allows researchers to obtain two time-frequency representations of event-related EEG responses \citep{tallon-baudry_oscillatory_1999}: 
\begin{itemize}
\item Evoked cortical activities are characterized by precise time-locking to the stimulus onset. They are obtained by 
estimating the wavelet power spectrum of the signal that results from the average of all the $N$ single EEG epochs $E_i$. To assess significant changes in the TF spectrum, the baseline correction is applied on the final wavelet transform.

\item In contrast to the evoked response, induced activities are characterized by a temporal shift from epoch to epoch \citep{grandchamp_single-trial_2011}. No phase relationship is assumed between the oscillatory responses and the stimulus onset. Hence, these activities cannot be revealed by classical averaging techniques. To analyse these induced activities, the wavelet power spectrum is firstly obtained for each $E_i$, then averaged. Following \citep{grandchamp_single-trial_2011}, significance test can be obtained by applying a baseline correction for each $E_i$ before the averaging. 
\end{itemize}

\paragraph{Normalization of TF maps}
To normalize the resulting TF spectrum by a baseline, four approaches are available:
\begin{itemize}
\item Relative baseline: the mean and standard deviation are calculated inside the baseline for each frequency band in the TF map. Such values are used to normalize all time points at each frequency to provide a time-frequency map in standard deviations (a \textit{z-score}) of the values observed during the baseline.
\item Subtractive baseline: consists in removing the mean of the baseline from each frequency band of the TF map \citep{grandchamp_single-trial_2011}.
\item Divisive baseline: normalized TF maps are obtained by the ratio between the wavelet power spectrum and the mean of the baseline obtained at each frequency.
\item Decibel baseline: is the result of ten times the log-transformed divisive baseline correction \citep{grandchamp_single-trial_2011}. Changes in the TF maps are expressed in decibels (dB).
\end{itemize}

\paragraph{Baseline correction based on permutation methods to assess significant differences}
To find the significant time-frequency points with respect to the chosen baseline, we use a permutation (non-parametric) method inspired from \citep{grandchamp_single-trial_2011}. Briefly, this method consists in randomly selecting a baseline period within the epochs limits and then applying them to other (different) randomly selected epochs. 
We repeat this procedure $Np$ = 200 times to have the same number of time-frequency surrogate distributions and hence obtain the histograms of the mean values across time for each frequency bin. Then, values on the original induced time-frequency map lying outside the 100*$\alpha$ \% of the tails are considered as statistically different with respect to the baseline at a significant level $p < \alpha$. To take into account the multiple comparisons problem, $\alpha$ is divided by the number of time samples according to the Bonferroni correction.

For evoked activity, since time-frequency maps are not computed by epoch $E_i$, surrogate distributions are obtained as follows:  First, $Np$ subsets containing $N$/2 epochs randomly selected are formed. Each subset is averaged, then transformed to an evoked time-frequency map that is baseline-corrected. As the permutation approach above, baseline period is placed randomly within an epoch then permuted with that from another evoked TF map.

%%%%%%%%%%%%%%%%%%%%%%% APPLICATIONS %%%%%%%%%%%%%%%%%%%%%%%%
\section{Case studies using real data}
To illustrate the utility of the proposed toolbox, we used data from two different databases that include EEG and cardio-respiratory recordings. We firstly used CARE-rCortex to compare time-frequency maps from EEGs in two different respiratory conditions: normal breathing \textit{versus} forced breathing. In a second example, we used a polysomnography recording from the Physionet database to compare heartbeat activity at different sleep stages. Likewise, each case use was implemented using a different event-related approach: evoked activity to compare respiratory conditions, and induced activity to assess heartbeat potentials. 

\begin{figure*}[t!]
\centering %from left, bottom, right and top
\includegraphics[width= \textwidth]{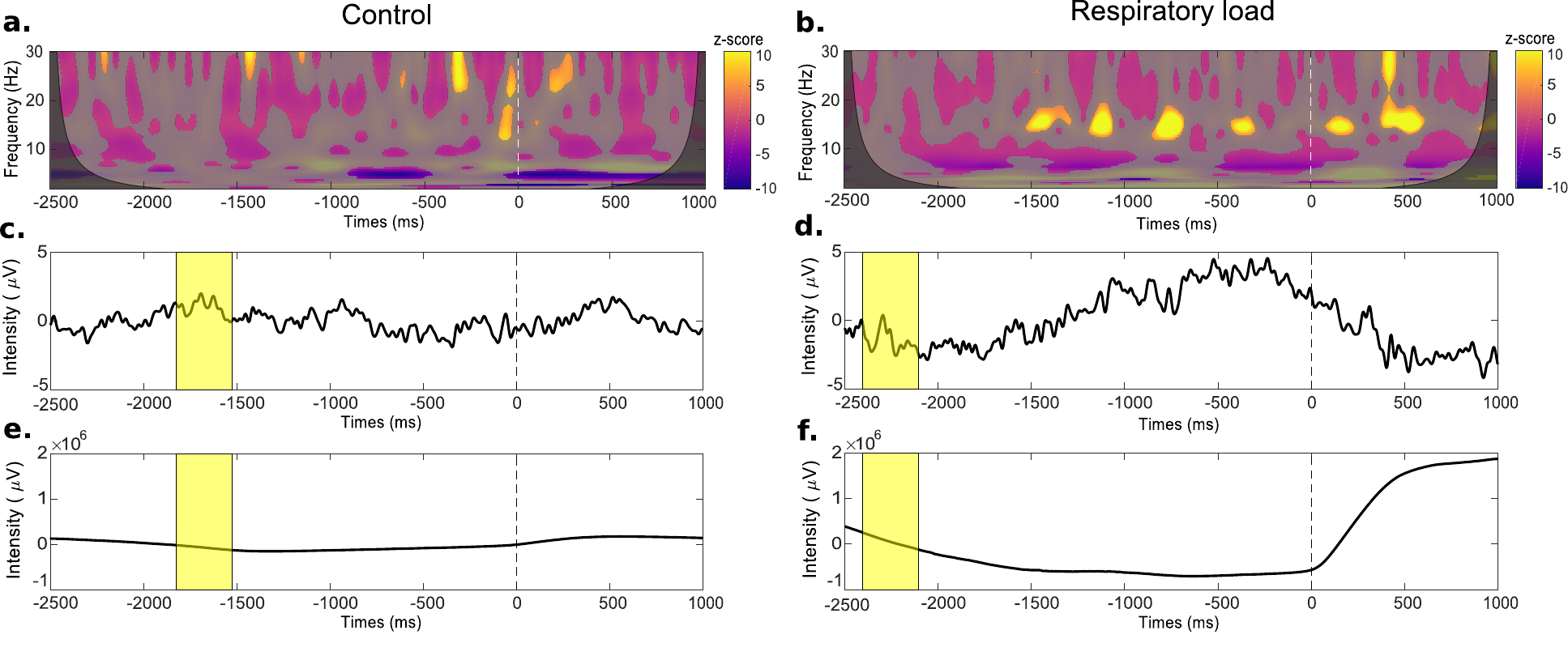}
\caption{CARE-rCortex results of respiratory event-related EEG potentials under control \textit{versus} inpiratory load conditions. TF maps of Cz activities under the control condition \textit{(a)} and the load condition \textit{(b)} for a healthy subject. The non-significant mask with respect to the baseline is superimposed to each of the time-frequency map. Cone of influence is displayed on each of the time-frequency map as a superimposition of the time-frequency map and its significant mask. EEG activity averaged across the epochs of Cz in each condition (control \textit{(c)} \textit{versus} charge \textit{(d)}) with the median baseline in yellow. Averaged pressure across the epochs in each condition (control \textit{(e)} \textit{versus} inspiratory load \textit{(f)}) with the median baseline indicated by the yellow rectangles.}
\label{app_respi}
\end{figure*}

\subsection{Respiratory analysis}
\paragraph{Protocol and data}
This study was carried out on healthy subjects breathing through a mouthpiece (see \citep{hudson_2016} for a detailed description of the protocol, ethical approval and recording devices). EEG signals were recorded by surface electrodes along with the breathing activity using a pneumotachograph at 2500 Hz sampling rate. We selected respiratory pressure and Cz channel data from one subject during normal breathing (10 minutes) and inspiratory load (10 minutes). The experiment was designed to simulate breathing discomfort in ventilated patients when the mechanical ventilator is improperly adjusted. Breathing discomfort implies brain state changes that can be detected and studied by analysing the EEG activity \citep{navarro_bvi_2016}.

\paragraph{Data processing}
After importing data (the installation of \textit{BVA import/export} EEGLAB plug-in \citep{andreas_bva_2013} is required), it was processed according to the method described by Raux et al. \citep{raux_electroencephalographic_2007} and was down-sampled to 500Hz and low-pass filtered at 30Hz cut-off frequency. Respiratory marks were automatically detected by CARE-rCortex. EEG was segmented into 3.5 seconds excerpts according to the respiratory marks so the pre- and post-inspiratory times were 2.5 and 1 seconds respectively.
Then, the time-frequency analysis was done using CARE-rCortex. Because we are interested in pre-inspiratory activities, the baseline period, of 300 ms, was applied at the start of the expiration. We studied frequencies between 2 and 30Hz with a resolution of 0.25Hz. Regions in the TF map were detected as statistically different from the baseline activity with a $p<0.01$.

\paragraph{Results}
The results concerning respiratory-related activity can be seen in Fig. \ref{app_respi}-a and \ref{app_respi}-b. Significant regions in the time-frequency plane with respect to the baseline resulted in increased z-values during inspiratory load. Indeed, the existence of a respiratory pre-motor activity in this condition was especially noticeable around 15Hz between 1500 and 500 ms before inspiratory triggers. As expected, differences  between the two conditions (normal and forced breathing) were also evidenced in time domain averaged potentials (see Fig \ref{app_respi}-c and \ref{app_respi}-d). On the other hand, the absence of this pre-motor activity during normal breathing can be explained by the fact that the automatic breathing is normally controlled by the brain stem. Hence, inspiratory related activity was not relevant in time-frequency maps.

\subsection{Cardiac analysis}
\paragraph{Protocol and data} We selected a subject (\emph{slp45}) from the MIT-BIH Polysomnographic Database (see \citep{ichimaru_1999}).
Data were acquired during sleep for evaluation of chronic obstructive sleep apnea syndrome, and included (among other monitoring signals) one EEG (C3-O1) and ECG signals sampled at 250 Hz, plus the annotated beat-by-beat and sleep stage scorings. We only analysed segments corresponding to wakefulness and sleep stage 3 (the deepest sleep) as they typically present different heartbeat-related potentials (see Fig. \ref{cardiac_potential}).

\paragraph{Data processing}
Firstly, a bandpass filter (cutoff frequencies 1-40 Hz) was applied on both EEG and ECG signals, and heart beat marks (corresponding to R peaks) were detected by using ``Detect physiological events'' tool of CARE-rCortex. Then, EEG was segmented into 1-second trials, from 200 milliseconds before to 800 milliseconds after each beat mark. To reduce a potential effect of cardiac field artefact, we subtracted from each trial, a template of cardiac artefact. Such template was obtained as the median EEG epoch estimated over adjacent cardiac cycles (given the detected beat marks). Baseline was placed between -200 ms to -50 ms before the R peak detection to normalize the data by a z-score correction. Time-frequency analysis was done between 2 and 30~Hz with a frequency resolution of 0.25~Hz.  

\begin{figure*}[ht!]
\centering %from left, bottom, right and top
\includegraphics[width= \textwidth]{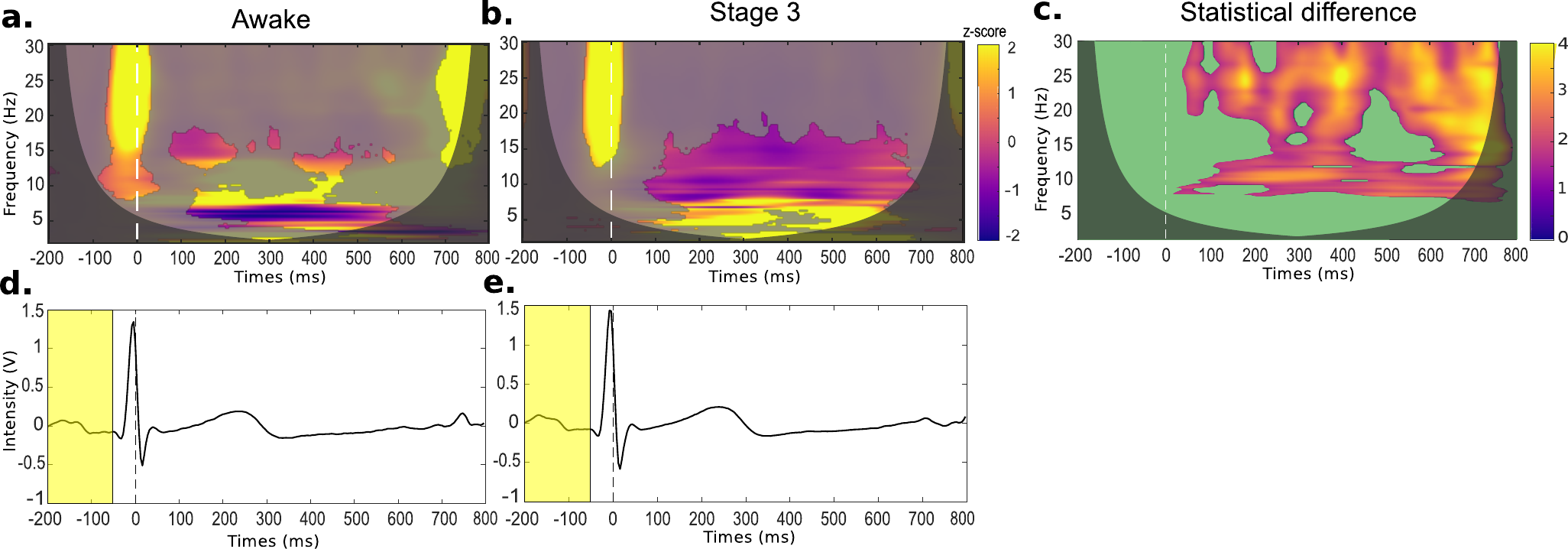}
\caption{Heartbeat event-related potentials during wakefulness and sleep stage 3 conditions for a healthy subject. 
Time-frequency maps of the EEG activity during wakefulness \textit{(a)} and the sleep stage 3 \textit{(b)}. Unmasked color regions in time-frequency maps indicate the significant activation levels with respect to the baseline ($p<0.01$).  The statistical differences between \textit{(a)} and \textit{(b)} are shown in \textit{(c)}, where the superimposed green mask highlights the significant differences between the two conditions. Black zones correspond to the cones of influence, i.e. regions not influenced by edge effects. Average of cardiac signals across the epochs with the median baseline (yellow rectangle), during wakefulness \textit{(d)} and sleep stage 3 \textit{(e)}.}
\label{app_cardiac}
\end{figure*}

\paragraph{Results}
Heartbeat induced activities in the EEG are depicted in Fig.~\ref{app_cardiac}-a and~\ref{app_cardiac}-b. Significant areas in the time-frequency map with respect to the baseline can be observed both in wakefulness and stage 3 sleep. Compared to wakefulness, an increase of slow (2-7 Hz) activities is clearly observed during sleep stage. Due to physiologically induced changes in heart rate~\citep{Snisarenko_1978}, an artefact template cannot be accurately estimate. This can explain that, despite the artefact reduction procedure, the large potential associated to the R-wave can still be clearly distinguished from baseline in both conditions.

Since each induced map is the average of an ensemble of time-frequency maps, we could also compare the differences between wakefulness and stage 3 using a permutation test~\citep{maris_2007}. In Fig. \ref{app_cardiac}-c we can observe that differences are mainly associated to alpha (around f $=10$~Hz) and faster activities (f $>$ 20~Hz) between 150 and 400 ms, and beyond t = 600~ms. Although differences between 150 and 400~ms could be due to differences in the propagation of ECG T-wave potentials \citep{dirlich_cfa_1997}, note that cardiac activities in that period are similar in both conditions, as shown in Figs. \ref{app_cardiac}-d-e. Differences in neural responses  can therefore be explained by the slow brain activities that predominate in deep sleep stages, and not by differences in heart activities. We note, however, that differences localized beyond t = 650~ms might be due to differences in cardiac activities recorded at the end of EEG epochs.

%%%%%%%%%%%%%%%% DISCUSSION and CONCLUSION %%%%%%%%%%%%%%%%%
\section{Discussion and conclusion}
In this paper, we have introduced an open source toolbox as an EEGLAB plug-in for EEG analysis in time-frequency domain. Contrarily to other tools that also offer TF analysis - like EEGLAB \citep{delorme_eeglab:_2004}, FieldTrip \citep{oostenveld_fieldtrip:_2010} or ERPWAVELAB \citep{morup_erpwavelab:_2007} - CARE-rCortex provides an easy-to-use interface to normalize TF maps with a baseline time-locked to cardio-respiratory events. Furthermore, CARE-rCortex allows the detection and the manual validation of CR events, necessary to define a baseline time-locked to a phase of the heart or breathing rhythms. A statistical test has also been integrated to highlight significant points of the TF maps with respect to the baseline.

To illustrate the functionalities of our plug-in, we have studied two real cardio-respiratory datasets. For the respiratory example, we have found a central pre-inspiratory activation of the EEG activity in inspiratory load condition, which is absent during spontaneous breathing. This finding corroborated the implication of central areas in the cortex to compensate breathing impairment \citep{dubois_neurophysiological_2016}. For the cardiac case, an increased synchronization (with respect to the baseline) in delta and theta activities was found during deep sleep. When comparing the awake and the sleep conditions, the main differences were observed in the high-alpha and the fast EEG bands, in particular around ECG T-wave, which cannot be explained by differences in cardiac activities. 

Brain activity changes during different cognitive or pathological brain states are involved in autonomic regulation. During these conditions, cardiac and respiratory autonomic modulation might partially depend on central nervous system modulation, allowing potential exploration of higher brain structure activity through peripheral autonomic modulation. The analysis of cardio-respiratory-related cortical activities could provide valuable information about brain states in respiratory or psychiatric disorders, sleep/dreaming stages, etc. 

Although the CARE-rCortex toolbox is designed to study EEG activities locked to heartbeats or breathing rhythms, the approach is applicable to any neuroimaging functional method (e.g. EEG, fMRI, and MEG signals) in both animal or human studies. One of the major limitations of the current toolbox concerns the detection of CR events which still needs a manual checking of the baseline locations. This issue could be improved in a further release. Another limitation results from artefacts recorded by the EEG electrodes. Before using CARE-rCortex, researchers can use existing plug-ins of EEGLAB to correct or reject EEG epochs contaminated with ocular or head movements, or by muscle artefacts. If several EEG channels are available, spatial filters or ICA-based methods can be applied to reduce the cardiac artefact. To complement the interface, other features could be implemented in the toolbox like phase synchrony or coherence computations. In future versions of the toolbox, we could also extend the event detection to other physiological events like the isotonic or isometric contractions in electromyographic signals.

The CARE-rCortex toolbox can be freely downloaded from \url{https://github.com/FannyGrosselin/CARE-rCortex} where a detailed tutorial explaining all the features of the toolbox is also available. The toolbox was developed with a Matlab version (R2017b) and the last release of EEGLAB (eeglab\_14\_1\_1b). The toolbox works on all major operating systems and is compatible with different versions of Matlab newer than R2013a.   

%%%%%%%%%%%%%%%%%%%% ACKNOWLEDGMENTS %%%%%%%%%%%%%%%%%%%%%%
\section*{Acknowledgments}
The work of X. Navarro-Sune was supported by Air Liquide Medical Systems S.A., France.

%%%%%%%%%%%%%%%%%%%%%%% REFERENCES %%%%%%%%%%%%%%%%%%%%%%%%%
\section*{References}
%\bibliography{References}

\end{document}